\documentclass[aps,preprint]{revtex4}
\usepackage{graphicx}
\usepackage[dvips]{epsfig}
\begin{document}
\title{Infrared absorption and Raman scattering
on coupled plasmon--phonon modes
in  superlattices}
\author{L. A. Falkovsky$^{1}$, E. G. Mishchenko$^{1,2}$ }
%\email{falk@itp.ac.ru}
\affiliation{$^{1}$Landau Institute for Theoretical Physics,
119337
Moscow, Russia\\
%\author{Е.Ж. Мищенко}
%\affiliation{
$^{2}$Department of Physics, University of Utah, Salt Lake City,
UT 84112}

\begin{abstract}
We consider theoretically a  superlattice formed by  thin
conducting layers separated spatially between insulating layers.
The dispersion of two coupled phonon-plasmon modes of the system
is analyzed by using Maxwell's equations, with the influence of
retardation included. Both transmission for the finite plate  as
well as absorption  for the semi-infinite superlattice in the
infrared are calculated. Reflectance minima are determined by the
longitudinal and transverse phonon frequencies in the insulating
layers and by the density-state singularities of the coupled
modes. We evaluate also the Raman cross section from the
semi-infinite superlattice.

\end{abstract}
 \maketitle

\section{Introduction}

Coupling of collective electron oscillations (plasmons) to optical
phonons in polar semiconductors was predicted more than four
decades ago \cite{Y}, experimentally observed using Raman
spectroscopy in $n$-doped GaAs \cite{MW} and extensively
investigated  since then (see, e.g., \cite{IP}). Contrary, the
interaction of optical phonons with plasmons in the semiconductor
superlattices is much less studied. A two-dimensional electron gas
(2DEG) created at the interface of two semiconductors has
properties which differ drastically from the properties of its
three-dimensional counterpart. In particular, the plasmon spectrum
of the 2DEG is gapless \cite{St} owing to the long-range nature of
the Coulomb interaction of carriers, $\omega^2(k)=v_F^2 \kappa_0
k/2$, where $v_F$ is the Fermi velocity and $\kappa_0$ is the
inverse static screening length in the 2DEG.  Coupling of
two-dimensional plasmons to optical phonons has been considered in
Refs. \cite{PWD,SH} for a single 2DEG layer. The resulting
coupling is usually non-resonant since characteristic phonon
energies $\sim 30-50$ meV are several times larger than typical
plasmon energies. Still, hybrid plasmon--optical-phonon modes are
of considerable interest in relation  to polaronic transport
phenomena \cite{FMR}, Raman spectroscopy and infrared optical
absorption experiments.

Plasmon excitations in a periodic system of the electron layers
have been discussed in a number of theoretical papers
\cite{F,Quinn}, disregarding the phonon modes. In the present
paper we analyze the coupled phonon-plasmon modes for a
superlattice of 2D electron layers sandwiched between insulating
layers and demonstrate a possibility of stronger resonant coupling
of plasmons to optical phonons excited in the insulator. This
enhancement occurs in superlattices due to interaction of plasmons
in different layers which spreads plasmon spectrum into a
mini-band spanning the energies from zero up to the new
characteristic energy, $v_F(\kappa_0/d)^{1/2}$, where $d$ is the
interlayer distance \cite{MA}. This value could exceed typical
phonon frequencies leading to formation of resonant hybrid modes
around crossings of phonons and band plasmons.

Usually the coupled phonon-plasmon modes are considered in the
so-called electrostatic approximation when the retardation effect
is ignored and the terms of  $\omega/c$ in the Maxwell equations
are neglected in comparison with the terms having values of the
wave vector $k$. This is correct if  the Raman scattering is
studied, when $\omega$ has a meaning of the frequency transfer.
Then, $\omega$ is much less then the incident frequency  of the
order of $\omega^i\sim ck$. But if we interested in absorption for
the infrared region, when $\omega$ is the frequency of incident
light and corresponds to the optical phonon frequency, $\omega$
and $ck$ are comparable. In this case, the retardation effect must
be fully included.

The plan of the present paper  (preliminary results were published
in Ref. \cite{FM}) is the following. In Sec. II, the Maxwell
equations for the periodic system of thin conducting layers
sandwiched between the insulating layers are solved to find a
spectrum of coupled phonon-plasmon modes. In Sec. III, we consider
absorption of the finite sample of the layers and reflectance of
the semi-infinite system  in the infrared region. In Sec. IV, we
analyze the  Raman light scattering from the semi-infinite system
of layers.

\section{Spectrum of coupled modes}

\begin{figure}[h]
\resizebox{.50\textwidth}{!}{\includegraphics{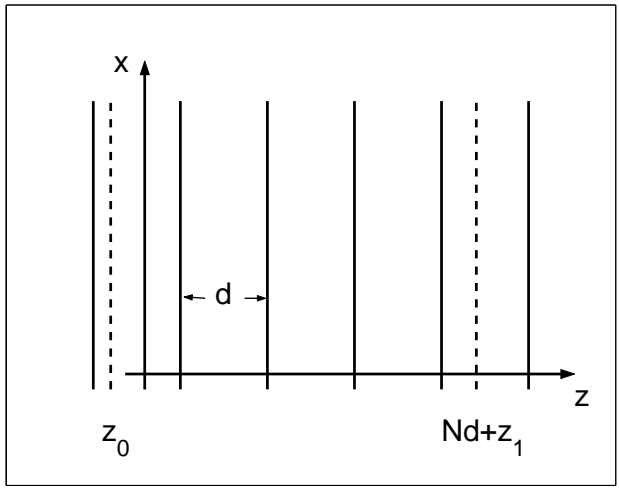}}
\caption{\label{stl} The  stack of 2d conducting layers separated
between dielectric layers of thickness $d$ with dielectric
constant $\varepsilon (\omega)$; $z_0$ and $Nd+z_1$ are the
boundaries (dashed lines) of the sample if the finite stack is
considered.}
\end{figure}

We consider a superlattice formed by periodically grown layers of
two polar  semiconductors ( e.g.,\, GaAs and AlGaAs) with 2DEG
layers formed in the interface regions, see Fig.~\ref{stl}.

For the sake of simplicity we assume a superlattice with a single
period $d$ and the thickness of a 2DEG layer much less than the
period. We also neglect the difference in bulk phonon properties
of the two materials. Optical phonons in polar semiconductors are
most conveniently described within the dielectric continuum model,
which yields the familiar phonon contribution into the dielectric
function,
$$\varepsilon(\omega)=\varepsilon_{\infty} \frac{\omega^2_{\rm
LO}-\omega^2-i\omega\Gamma}{\omega^2_{\rm
TO}-\omega^2-i\omega\Gamma}, $$ where $\omega_{\rm LO}$ and
$\omega_{\rm TO}$ are the frequencies of longitudinal and
transverse optical phonons, respectively, and $\Gamma$ is the the
phonon width (we do not distinguish the widths of the TO and LO
modes).

Collective modes of our system are described by the Maxwell
equations, which in the Fourier representation with respect to
time have the form
\begin{equation}
\label{maxwell} \nabla (\nabla \cdot {\bf E}) -\nabla^2 {\bf
E}=\varepsilon\frac{\omega^2}{c^2}{\bf E}+ \frac{4\pi
i\omega}{c^2}{\bf j},
\end{equation}
where the last term takes into account the in-plane electric
currents ${\bf j}$ induced in the 2DEG layers by the electric
field ${\bf E}$. As usual, when the frequency of collective mode
lies above the electron-hole continuum, it is sufficient to use
the Drude conductivity to describe the in-plane electric currents,
\begin{equation}
\label{current}{\bf j}_{\parallel}(\omega,x,z)=
\frac{ie^2n_e}{m(\omega+i\gamma)}\sum_{n}\delta(z-z_n) {\bf
E}_{\parallel}(\omega,x,z),
\end{equation}
where $z_n=nd$ are positions of interfaces ($n$ is integer
corresponding to the periodicity of the stack),
 $n_e=p_F^2/2\pi \hbar^2$ and $m$ are the electron density (per
 the  surface unit) and the effective mass,
 $\gamma$ is the electron collision frequency,  $x$ and
$z$ are the coordinates along and perpendicular to the interfaces,
respectively.

We consider the case of the $p-$polarization, when the field ${\bf
E}$ lies in the $xz-$plane and, therefore, the current ${\bf j}$
has then only the $x-$component. Making use of the Fourier
transformations with respect to the $x-$coordinate, ${\bf E}\sim
e^{i k_xx}$,  we can rewrite the Maxwell equations (\ref{maxwell})
in the form
\begin{eqnarray}\nonumber
ik_x\frac{dE_z}{dz}-\frac{d^2E_x}{dz^2}-\varepsilon(\omega/c)^2E_x
=\frac{4\pi i\omega}{c^2}j_x\\ \nonumber
ik_x\frac{dE_x}{dz}+(k_x^2-\varepsilon\omega^2/c^2)E_z=0.
\end{eqnarray}
Eliminating
$$ E_z=-\frac{ik_x}{\kappa^2}\frac{dE_x}{dz},$$
we get  for $E_x$ the equation
\begin{equation}\label{eqs}
\left(\frac{d^2}{dz^2}-\kappa^2-2\kappa
C\sum_{n}\delta(z-z_n)\right)E_x(\omega,k_x,z)=0,
\end{equation}
where $$C=\frac{2\pi n_e e^2\kappa}{\varepsilon
\omega(\omega+i\gamma)m},\quad
\kappa=\sqrt{k_x^2-\varepsilon\omega^2/c^2}.$$
 At the interfaces
 $z=z_n$, the $E_x$ component must be continuous and
  the $z-$component of  electric induction $\varepsilon E_z$ has a jump:
 \begin{equation}\label{jump}
\varepsilon ( E_z|_{z=nd+0}- E_z|_{z=nd-0})=4\pi
\int_{nd-0}^{nd+0} \rho(\omega,k_x,z)dz,
\end{equation}
where the carrier density is connected to the current
(\ref{current}) according to the equation of continuity
\begin{equation}\label{dens}
\rho(\omega,k_x,z)=j_x(\omega,k_x,z)k_x/\omega.
\end{equation}
%We have to construct the solutions for our periodic superlattice
%in the Bloch form. It is convenient to choose the solutions to Eq.
%(\ref{eqs}) vanishing at the interval boundary, for instance, at 0
%and $d_1$:

For the infinite stack of layers, $-\infty < n<\infty$,
independent solutions to Eqs.~(\ref{eqs}) -- (\ref{jump})
represent two Bloch states $E_x(z)=f_{\pm}(z)$ moving along
positive and negative directions of the $z$-axis,
\begin{equation}\label{ef}
f_{\pm}(z)=e^{\pm ik_znd}{\sinh{\kappa(z-nd)}-e^{\mp ik_zd}
\sinh{\kappa[z-(n+1)d]}},\quad
 nd<z<(n+1)d
\end{equation}
with  the quasi-momentum $k_z$  determined from
%the equations for the coefficients in
%Eq. (\ref{ef}) and
 the dispersion equation:
\begin{equation}\label{diseq}
\cos{k_zd}=\cosh{\kappa d}-C\sinh{\kappa d}.
\end{equation}

The quasi-momentum $k_z$ can be restricted to the  Brillouin
half-zone $0<k_z<\pi/d$, if the phonon and electron damping
vanishes. In the general case, we fix the choice of the
eigenfunctions in Eq.~(\ref{ef}) by the condition Im$k_z>0$, so
that the solution $f_{+}$ decreases in the positive direction $z$.
\begin{figure}[h]
\resizebox{.40\textwidth}{!}{\includegraphics{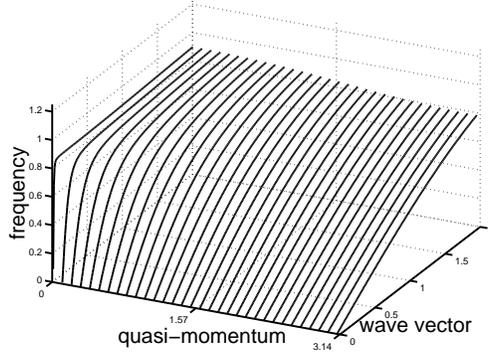}}
\caption{\label{p34} The  plasmon-like (at small wave vectors)
mode of an infinite number of 2D metallic layers (with the carrier
concentration in every layer  $n_e=3\cdot10^{11}$ cm$^{-2}$)
sandwiched between dielectric layers of thickness $d$ with the
phonon frequency $\omega_{\rm LO}$. The frequency
$\omega_{-}(k_x,k_z)$ in units of $\omega_{\rm LO}$ is plotted as
a function of the in-plane wave vector $k_x$  and the
quasi-momentum $k_z<\pi/d$ (both in units $1/d$).  Parameters used
here are reported in the literature for GaAs: $\omega_{\rm
LO}=36.5$ meV, $\omega_{\rm TO}=33.6$ meV,
$\varepsilon_{\infty}$=10.6; the thickness $d$ is taken as
$d=1/\kappa_0$,
$\kappa_0=2me^2/\hbar^2\varepsilon_{\infty}=2.5\cdot10^6$
cm$^{-1}$ is the screening length.}
\end{figure}
\begin{figure}[h]
\resizebox{.40\textwidth}{!}{\includegraphics{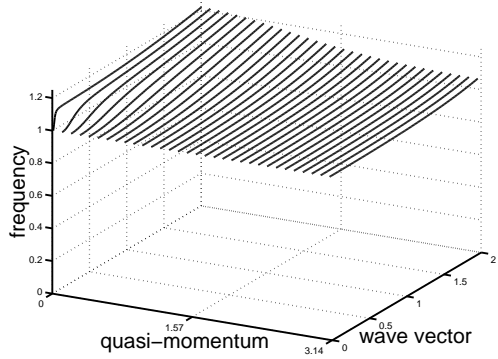}}
\caption{\label{p35} Same as in Fig. \ref{p34} but for the
phonon-like (at small wave vectors) mode $\omega_{+}(k_x,k_z)$.}
\end{figure}
The equation (\ref{diseq}) implicitly determines the spectrum
$\omega_{\pm}(k_x,k_z)$ of two coupled plasmon--optical-phonon
modes shown in Figs. \ref{p34}-\ref{p35} as  functions of the
in-plane wave vector $k_x$ and the quasi-momentum $k_z$. These
modes are undamped if  the electron collision rate and the phonon
width are small. The modes arise from the interaction of the
plasmon branch in the 2DEG and the phonon LO mode in the 3-d
insulator. They have a definite character far from the
intersection of the corresponding dispersion curves. For instance,
at small values of $k_x$, the $\omega_{-}(k_x,k_z)$ mode is mainly
the plasmon mode, whereas the $\omega_{+}(k_x,k_z)$ mode has
mainly the plasmon character. At large values of $k_x$, they
interchange their character.
%The spectrum is periodic in the mini-Brillouin zone,
%$\omega(k_x,k_z)=\omega(k_x,k_z+2\pi/d)$.

\begin{figure}[h]
\resizebox{.40\textwidth}{!}{\includegraphics{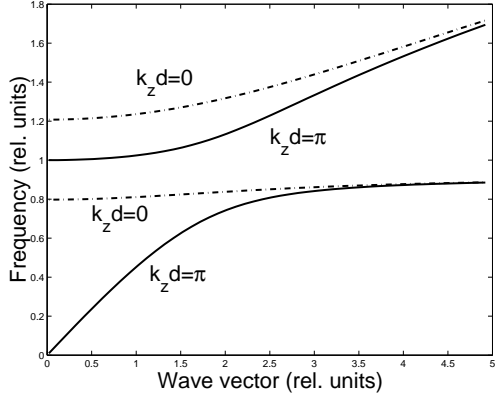}}
\caption{\label{phpl1} Dispersion of coupled phonon-plasmon modes
of an infinite number of 2D metallic layers sandwiched between
dielectric layers.   The frequencies $\omega_{\pm}(k_x,k_z)$ (in
units of $\omega_{\rm LO}$) are plotted as  functions of the
in-plane wave vector $k_x$ for two values of the quasi-momentum
$k_z$ (all in units of the inverse period $1/d=2.5\cdot10^6$
cm$^{-1}$). The dashed-dotted
 lines represent the upper boundary $\omega_{\pm}(k_x,k_z=0)$ of two
 phonon-plasmon modes, while the  solid lines mark the lower boundary
corresponding to $\omega_{\pm}(k_x,k_z=\pi/d) $. Parameters are
the same as in Figs. \ref{p34} - \ref{p35}.}
\end{figure}

\begin{figure}[h]
\resizebox{.40\textwidth}{!}{\includegraphics{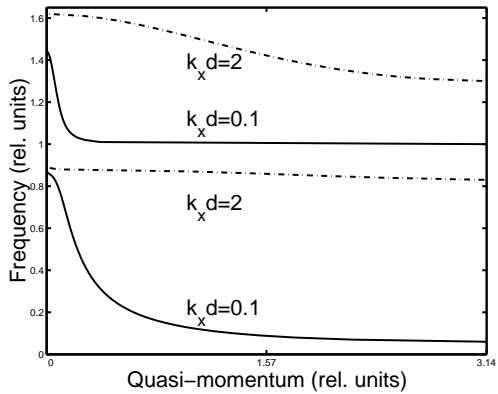}}
\caption{\label{pq1} Dispersion of coupled phonon-plasmon modes.
The frequencies $\omega_{\pm}(k_x,k_z)$  are plotted as a function
of the quasi-momentum $k_z$ for two values of the in-plane wave
vector $k_x$. }
\end{figure}

Note, that  $k_x=k_z=0$ is a saddle point for both branches
$\omega_{\pm}(k_x,k_z)$. In the vicinity of this point, the
frequency grows as a function of $k_x$ and decreases with
increasing $k_z$. This is evident in Figs. \ref{phpl1}-\ref{pq1}
and can be explicitly shown in the limit of $k_x \gg \omega
|\varepsilon (\omega)|/c$, when the retardation effects of the
electromagnetic field are negligible. Then, $\kappa\approx k_x$
and equation (\ref{diseq}) yields two solutions,
\begin{eqnarray}\label{der}
\omega^2_{\pm}(k_x,k_z)=\frac{1}{2}(\Omega^2+\omega_{\rm LO}^2)
\pm
  \frac{1}{2}\Bigl[(\Omega^2+\omega_{\rm LO}^2)^2-
  4\Omega^2\omega_{\rm TO}^2\Bigr]^{1/2},
\end{eqnarray}
where we introduced the notations,
%\begin{equation} \nonumber
$$\Omega^2(k_x,k_z)=\omega_0^2(k_x)
\frac{\sinh{k_xd}}{\cosh{k_xd}-\cos{k_zd}},\quad
\omega_0^2(k_x)=\frac{2\pi n_ee^2 k_x}{m\varepsilon_{\infty}}.$$
%\end{equation}
The frequency $\omega_0(k_x)$ is the conventional square-root
spectrum of two-dimensional plasmons in the limit of layer
separation large compared to the wavelength, $d \gg k^{-1}_x$. The
quantity $\Omega(k_x,k_z)$ describes the plasmon spectrum that
would exist in a superlattice of consisting non-polar
semiconductors (when $\omega_{\rm LO} =\omega_{\rm TO}$).
% The coupling constant
%$\lambda=\sqrt{\kappa_0 v^2/\varepsilon_{\infty} d}/\omega_{\rm
%LO}$
%determines the interaction between the plasmon and phonon
%modes.
%\begin{figure}[h]
%\resizebox{.40\textwidth}{!}{\includegraphics{pl2.eps}}
%\caption{\label{phpl2} Same as in Fig. (\ref{phpl1}) but for
%$\lambda=0.7$.}
%\end{figure}

  The mode $\omega_+(k_x,k_z)$ has a
gap while the other mode
 has a linear dispersion $\omega_{-}(k_x,k_z)=s(k_z)|k_x|$
at $k_x \ll (d^{-1}, k_z)$ with the velocity given by
$$ s(k_z)=v_F\sqrt{\frac{\kappa_0 d}{2(1-\cos{k_zd})}},$$
where $\kappa_0 =2me^2/\hbar^2\varepsilon_{\infty}$ is the static
screening radius in the 2DEG. For a fixed value of the in-plane
wave vector $k_x$, every mode develops a band (with respect to the
quasi-momentum $k_z$) with the boundaries,
 $\omega_{upper}(k_x) \le \omega (k_x,k_z) \le \omega_{lower}(k_x)$
 where
 $$\omega_{upper}^2=\Omega^2(k_x,0)=\omega_0^2(k_x)
\coth{\frac{k_x d}{2}}$$
%\end{equation}
for the upper boundary and
%\begin{equation}
 $$\omega_{lower}^2=\Omega^2(k_x,\pi/d)=\omega_0^2(k_x)
\tanh{\frac{k_x d}{2}}$$
%\end{equation}
for the lower boundary.  Fig.~\ref{pq1} illustrates the behavior
of the gapless, so called "acoustic"\, mode
 for the small values of $k_z<k_x$, where it acquires a gap.
 We see that the frequency of this mode
  decreases  rapidly in the region $k_z>k_x$.

In the rest of the paper we analyze various experimental
implications resulting from the existence of hybrid plasmon-phonon
modes. Such modes can be observed in both the infrared absorption
and the Raman spectroscopy.

\section{Infrared absorption on coupled plasmon--phonon modes}

 We calculate now the reflectance and the
transmission of a plane wave with the $p$-polarization, incident
from the vacuum on the thin plate consisting of the system of
layers. Let us suppose that the boundaries of sample are parallel
to the layers and intersect the $z-$axis at $z=z_0$  and
$z=Nd+z_1$, respectively, with $0<z_0, z_1<d$ (see Fig.
\ref{stl}).  We assume in the vacuum ($z<z_0$),
$$E_x(z)=e^{ik_z^{(i)}(z-z_0)}+Ae^{-ik_z^{(i)}(z-z_0)},$$
where $k_z^{(i)}=\sqrt{(\omega/c)^2-k^2_x}$ and $A$ is the
amplitude of the reflected wave. In the region $z>Nd+z_1$, the
transmitted wave has the form
$$E_x(z)=Te^{ik_z^{(i)}(z-z_1-Nd)},$$
and we search inside the plate the field as a sum of two solutions
(\ref{ef}):
$$E_x(z)=C_{+}f_{+}(z)+C_{-}f_{-}(z),$$
with the definite values of  $k_x$, $\omega$,  and
$k_z=k_z(\omega,\kappa)$ obeying the dispersion equation
(\ref{diseq}).
\begin{figure}[h]
\resizebox{.50\textwidth}{!}{\includegraphics{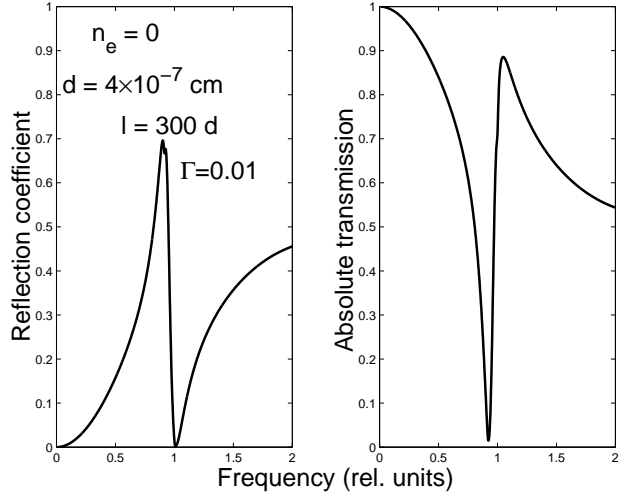}} \caption{
\label{0} -- Fig. \ref{123}:  Calculated p-polarized
reflection--absorption spectra of GaAs plates (of the thickness
$l=300d$) with the super-lattices of different electron
concentrations in a layer; the frequencies, the phonon width
$\Gamma$, and  the electron relaxation frequency $\gamma$ are
given in units of $\omega_{\rm LO}$. The lattice period
$d=1/\kappa_0$  and the incidence angle  $\theta=\pi/4.$
Parameters used here are reported in the literature for GaAs:
$\omega_{\rm LO}=36.5$ meV, $\omega_{\rm TO}=33.6$ meV,
$\varepsilon_{\infty}=10.6.$}
\end{figure}
\begin{figure}[h]
\resizebox{.50\textwidth}{!}{\includegraphics{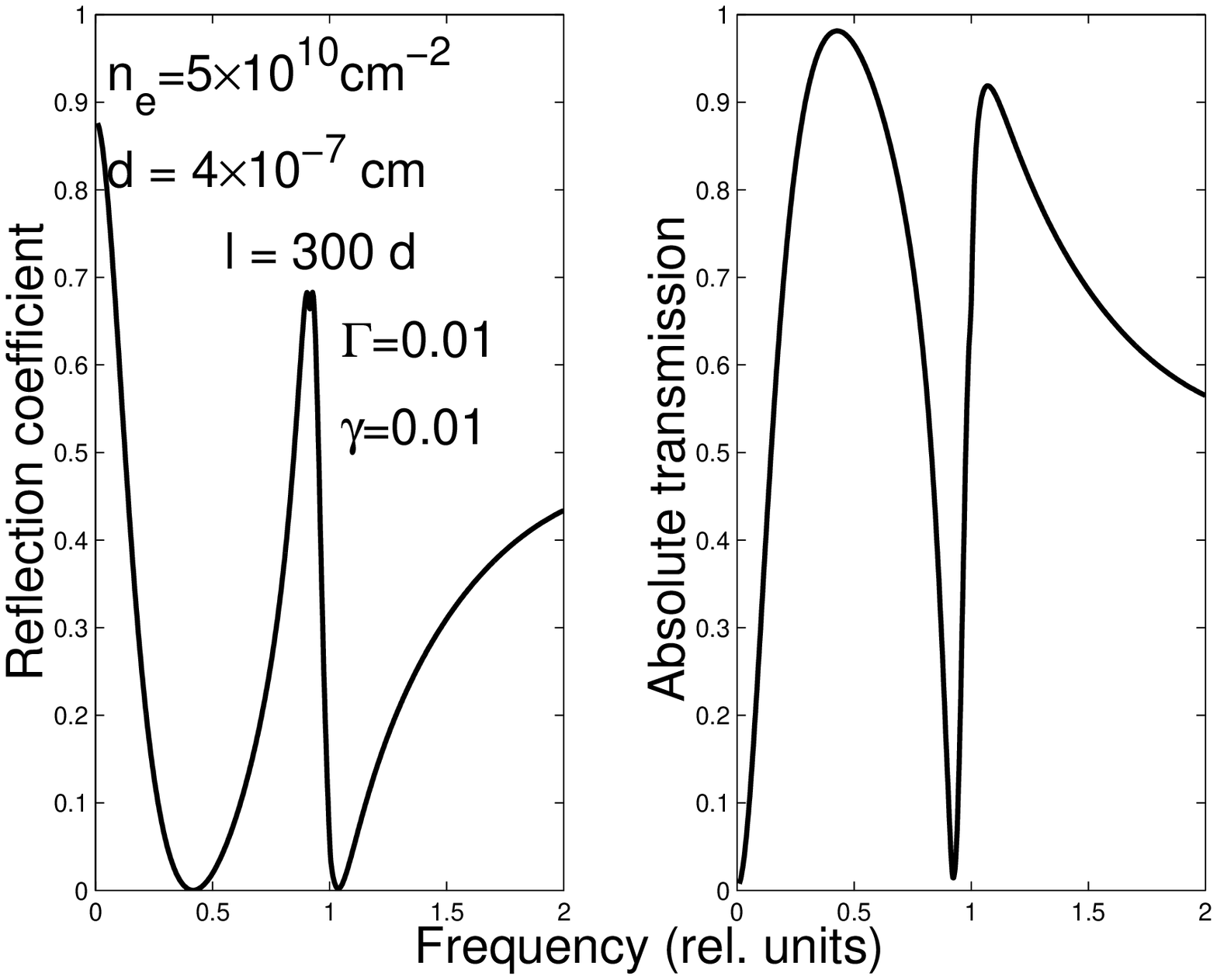}}
\caption{\label{53}  }
\end{figure}
%\begin{figure}[h]
%\resizebox{.50\textwidth}{!}{\includegraphics{retr23.eps}}
%\caption{\label{23}  }
%\end{figure}
\begin{figure}[h]
\resizebox{.50\textwidth}{!}{\includegraphics{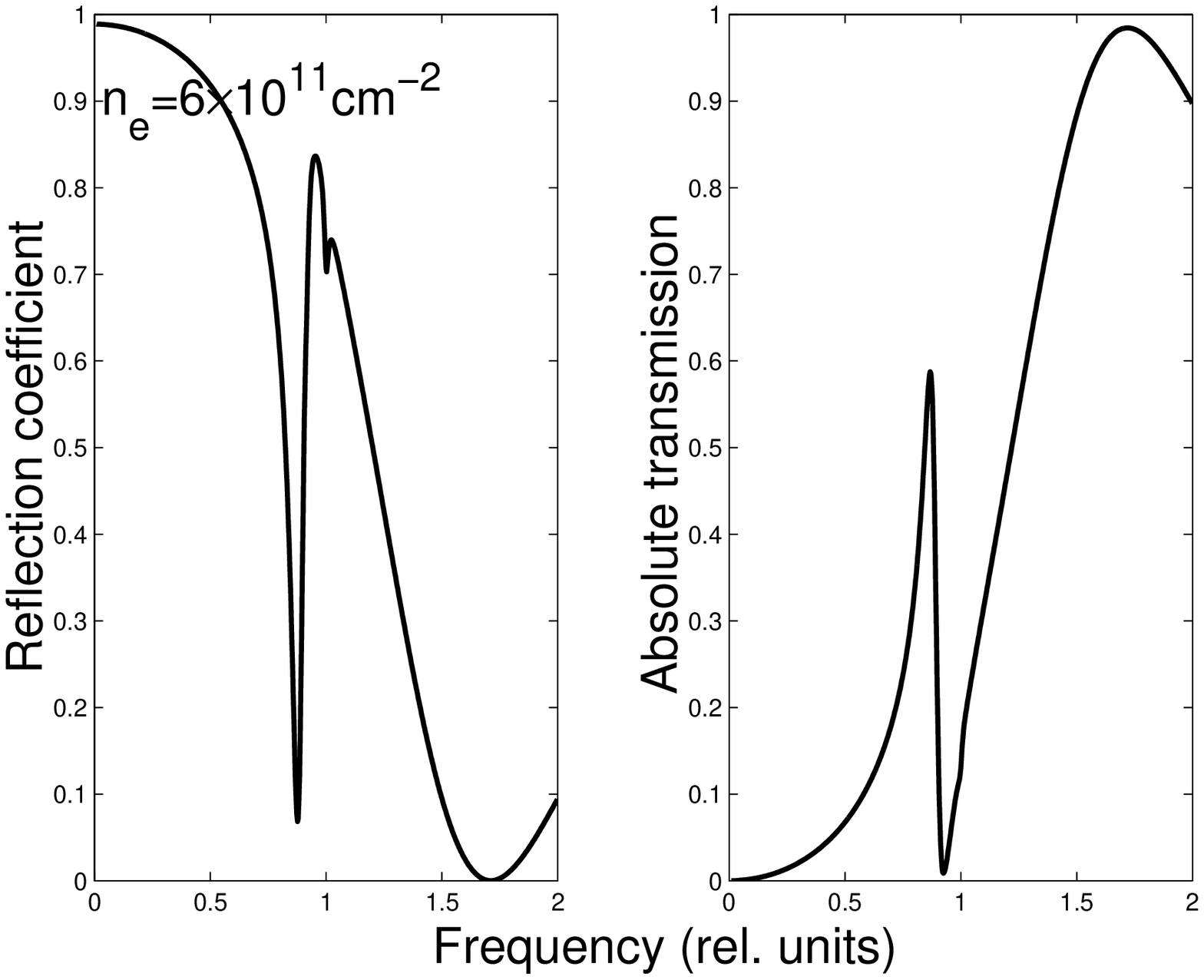}}
\caption{\label{63}  }
\end{figure}
%\begin{figure}[h]
%\resizebox{.50\textwidth}{!}{\includegraphics{retr83.eps}}
%\caption{\label{83}  }
%\end{figure}
\begin{figure}[h]
\resizebox{.50\textwidth}{!}{\includegraphics{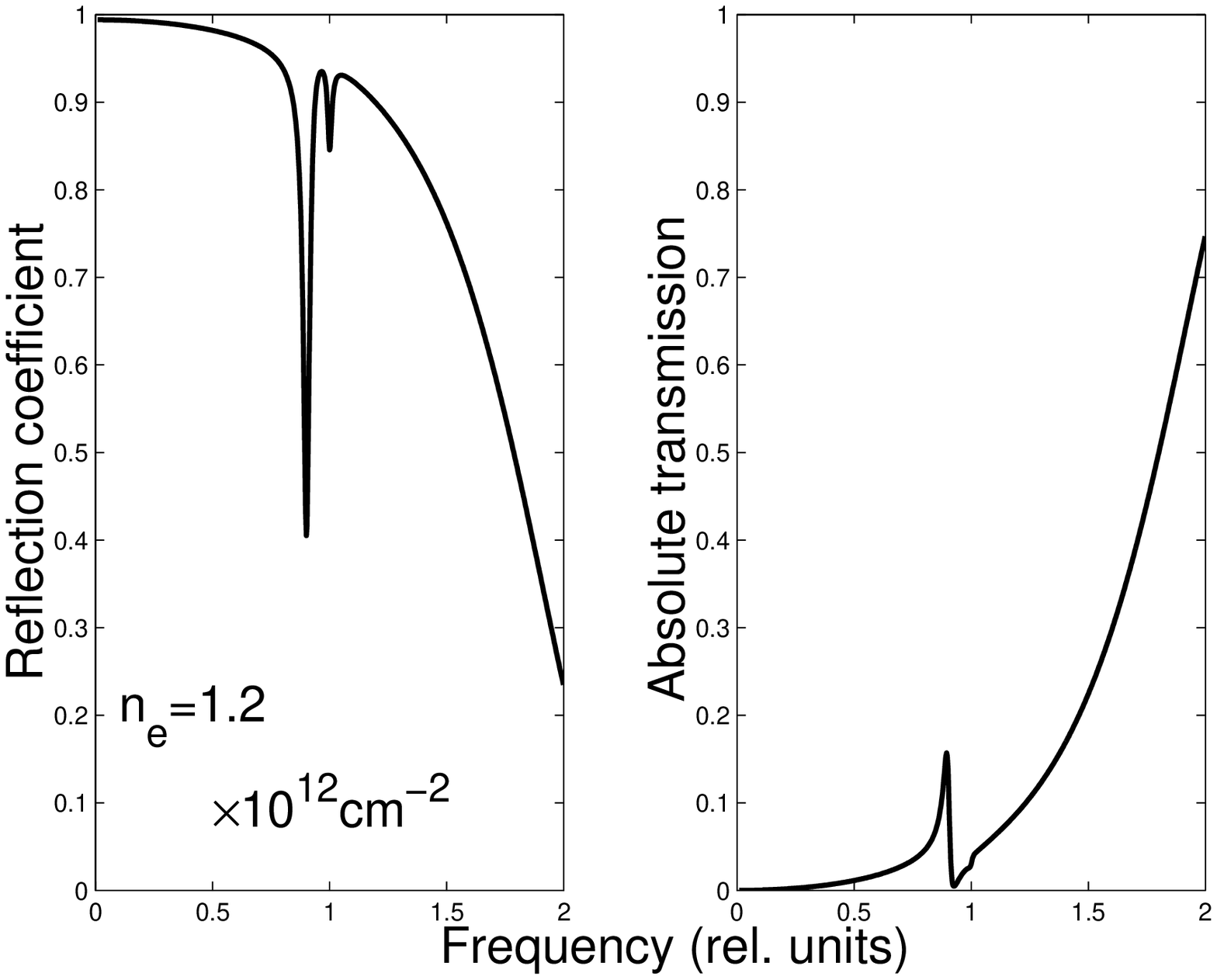}}
\caption{\label{123}  }
\end{figure}

 As usual, the
boundary conditions at $z=z_0$ and $z=Nd+z_1$ require the
continuance of the $x-$component of the electric field $E_x$
parallel to the layers and  of the $z-$component of the electric
induction $D_z$ normal to the layers. The second condition
rewritten via the electric field $E_x$ gives, e.g., at $z=z_0$
$$\frac{\varepsilon}{\kappa^2}E'_x(z_0+)=
-\frac{1}{k_z^{(i)2}}E'_x(z_0-).$$

%Transmission coefficient $T$ throw the sample of thickness $nd$.
The boundary conditions  give for $C_+, C_-, A,$ and $T$ the
following equations
$$ 1+A=C_{+}f_{+}(z_0)+C_{-}f_{-}(z_0),$$
$$ -1+A=\left[C_{+}f_{+}'(z_0)+C_{-}f_{-}'(z_0)
\right]\frac{\varepsilon k_z^{(i)}}{i\kappa^2},$$
$$T=C_{+}f_{+}(z_1)+C_{-}f_{-}(z_1),$$
$$-T=\left[C_{+}f_{+}'(z_1)+C_{-}f_{-}'(z_1)\right] \frac{\varepsilon
k_z^{(i)}}{i\kappa^2}.$$

 Solving these equations, we find the transmitted amplitude
$$T=\frac{2\varepsilon
k_z^{(i)}}{i\kappa^2}\frac{f_{+}'(z_1)f_{-}(z_1)-f_{+}(z_1)f_{-}'(z_1)}
{a_{11}a_{22}-a_{12}a_{21}}$$
 and the reflected amplitude
$$A=-1+2\frac{a_{21}f_{-}(z_0)-a_{22}f_{+}(z_0)}
{a_{11}a_{22}-a_{12}a_{21}},$$
 where
$$a_{11}=f_{+}'(z_0)\varepsilon k_z^{(i)}/i\kappa^2-f_{+}(z_0),
a_{22}=f_{-}'(z_1)\varepsilon k_z^{(i)}/i\kappa^2+f_{-}(z_1),$$
$$a_{12}=f_{-}'(z_0)\varepsilon k_z^{(i)}/i\kappa^2-f_{-}(z_0),
a_{21}=f_{+}'(z_1)\varepsilon k_z^{(i)}/i\kappa^2+f_{+}(z_1).$$
%$$f_{1,2}(z_0)=\sinh{(\kappa z_0)}- e^{\mp
%ik_zd}\sinh{[\kappa(z_0-d)]},$$
%$$ f_{1,2}(z_1)=e^{\pm  ik_znd}\{\sinh{(\kappa z_1)}- e^{\mp
%ik_zd}\sinh{[\kappa(z_1-d)]}\}.$$

\begin{figure}[h]
\resizebox{.50\textwidth}{!}{\includegraphics{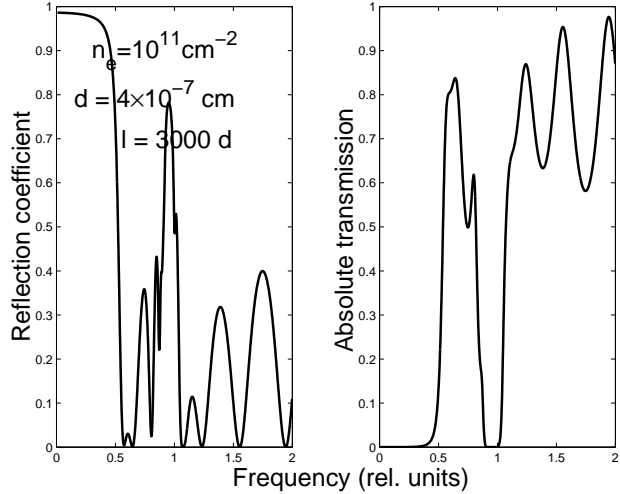}}
\caption{\label{rf3} Interference in the sample of thickness
$l=3000 d$; the electron concentration is $10^{11} $ cm$^{-2}$. }
\end{figure}

The transmission $|T|^2$ and reflection $|A|^2$ coefficients are
shown in Figs. \ref{0}-\ref{123} for  samples of various electron
concentrations $n_e$.  The incidence angle is taken
$\theta=\pi/4.$ Other parameters are the following: the thickness
$l$ = 300 $d$, the lattice period $d=1/\kappa_0=4\cdot10^{-7}$ cm,
the phonon width $\Gamma=0.02\omega_{\rm LO}$, and the electron
scattering rate $\gamma=0.01\omega_{\rm LO}$. In all that case
$k_z d<<1$, and there are many layers on the wave length
 of the field. Therefore, the reflectance is not sensitive
 to the positions of the sample surface $z_0$ and $z_1$. But, if
 the thickness $d$ is more larger, as shown in Fig. \ref{rf3},
 an interference phenomenon  is seen.

\begin{figure}[h]
\resizebox{.50\textwidth}{!}{\includegraphics{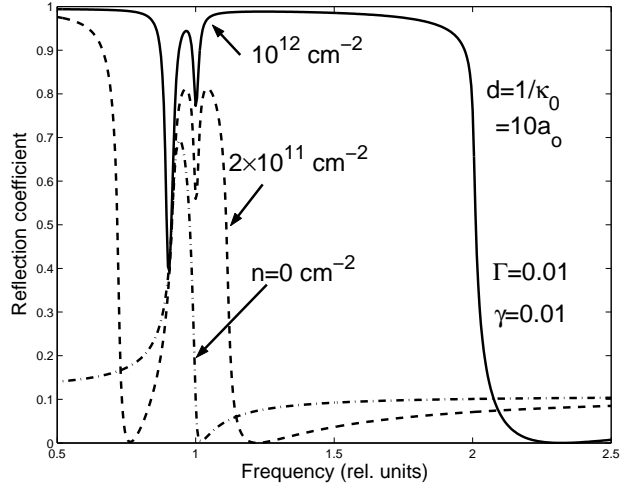}}
\caption{\label{rf1} Calculated p-polarized reflection-absorption
spectra for the semi-infinite superlattice via the frequency in
units of $\omega_{\rm LO}$ at the incident angle $\pi/4.$ The
electron concentration is labelled on curves, the lattice period
is $d=1/\kappa_0=4\times10^{-7}$ cm; the phonon width is $\Gamma$
 and the electron relaxation rate is
$\gamma$ (in uns. $\omega_{\rm LO}$). }
\end{figure}

\begin{figure}[h]
\resizebox{.50\textwidth}{!}{\includegraphics{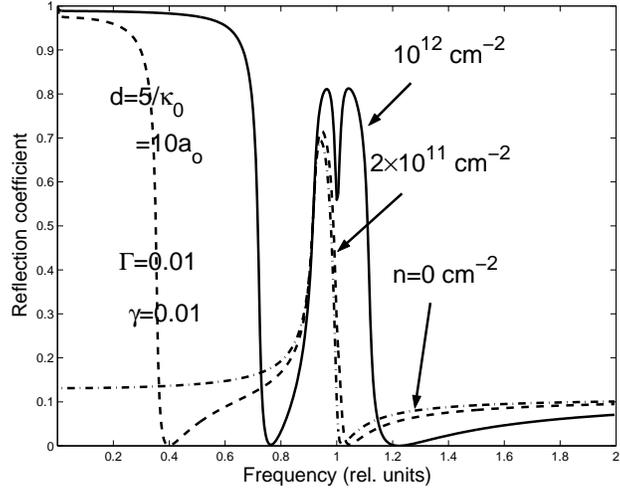}}
\caption{\label{rf2} Same as in Fig. \ref{rf1}, but for larger
period.}
\end{figure}

In order to avoid the interference effect, we calculate the
reflectance for the semi-infinite sample.  The results can be seen
in Figs. \ref{rf1} - \ref{rf2}, where the theoretical curves  are
presented for the various  electron concentrations and the lattice
period.
There is a singularity at  $\omega_{\rm LO}$. For $d=1/\kappa_0$
and the intermediate electron concentration in the layer (dashed
line in Figs. \ref{rf1}-\ref{rf2}), there are two regions (one is
at higher frequencies and another is just below $\omega_{\rm TO}$=
0.9$\omega_{\rm LO}$) where the sample with layers is more
transparent than the sample without any (shown in the
dashed-dotted line).

This is an effect of the coupled phonon-plasmon modes:  the minima
of the reflection coefficient correspond to the density state
singularities of the phonon-plasmon coupled modes at $k_z=0$ and
$k_x\rightarrow 0$ (see Fig. \ref{phpl1}). The frequencies of
these singularities are determined by Eq. (\ref{der}) with
$\Omega^2= 4\pi n_ee^2/md\varepsilon_{\infty}$. For the large
electron concentration (upper solid line), the reflection is not
complete only in the narrow interval bounded by the singularities
at $\omega_{\rm TO}$ and $\omega_{\rm LO}$. Finally, the
reflection coefficient tends to the unity at the low frequencies
because the skin depth of the metallic system goes to the infinity
in this case. For the sample with the large period $d=5/\kappa_0$
(Fig. \ref{rf2}), the effect of carriers is seen at the higher
concentration.

\section{Raman scattering from  coupled modes}

Consider now the Raman scattering of the radiation incident from a
vacuum with the vector potential ${\bf A}^i$, the frequency
$\omega^i$, and the wave vector ${\bf k}^i$ on the sample
occupying the semi-infinite space $z>z_0$, where $0<z_0<d$. The
corresponding quantities in the scattered wave are designated as
${\bf A}^s$, $\omega^s$, and ${\bf k}^s$.

In  addition to these fields, an  electric field ${\bf E}$ is
excited in the Raman light scattering in polar crystals along with
the longitudinal optical vibrations ${\bf u}$. The field ${\bf E}$
corresponds to the excitation of plasmons, whereas the vibrations
${\bf u}$ associate with the phonon excitations.

These processes can be described with the help of the effective
Hamiltonian
\begin{equation} \label{hami}
{\cal H} =
% {e^{2} \over m c^{2}}
 \int d^{3}r\,
 {\cal N}_{jk}(t,{\bf r})
 A^s_j(t,{\bf r})A^i_k(t,{\bf r}),
 \end{equation}
where the operator
\begin{equation} \label{ue}
{\cal N}_{jk}(t,{\bf r}) =
%\gamma\hat{ n}({\bf r},t)+
 g_{ijk}^u {\hat u}_i(t,{\bf r})+ g_{ijk}^E E_i(t,{\bf r})
\end{equation}
is linear in the phonon ${\bf u}$ and photon operators
 ${\bf E}$.

We are interested in the inelastic scattering on the
phonon-plasmon coupled modes. Therefore, we assume that the
frequency transfer $\omega=\omega^i-\omega^s$ is of the order of
the phonon frequencies $\omega_{\rm LO}$, but the frequencies
$\omega^i$ and $\omega^s$  of the incident and scattered fields
are much greater than the phonon frequencies. Then, we can ignore
the effect of carriers in the layers on the
 propagation of  both  scattered ${\bf A}^s(t,{\bf r})$ as well as  incident
 ${\bf A}^i(t,{\bf r})$ lights, taking the incident field
 in the sample ($z>z_0$) in the
form
%\begin{equation} \label{inc}
$$A^i_j(\omega^i,k^i_x,z)=
%\frac{2k_z^i}
%{k_z^i+\varepsilon(\omega^i)\omega^i\cos{\theta^i}/c}
t_je^{ik_z^i(z-z_0)},$$
%\end{equation}
 normalized to the incident flow;
 $t_j$ is the transmission coefficient from vacuum into the sample,
  $k_z^i=\sqrt{\varepsilon(\omega^i)(\omega^i/c)^2-k_x^{i2}}$.

When the scattered field ${\bf A}^s$ is taken as the variable in
the Hamiltonian (\ref{hami}), we obtain an additional current
${\bf j}^s(t,{\bf r})$
\begin{equation} \label{cs}
 -c\frac{\delta {\cal H}}{\delta
A^s_j}=j_j^s(t,{\bf r})=-
%\frac{e^2}{mc}
c{\cal N}_{jk}(t,{\bf r})A^i_k(t,{\bf r})
\end{equation}
in the Maxwell equation for this field. The phonon
 and plasmon fields in ${\cal N}_{jk}(t,{\bf r})$ are the source of the
 Raman scattering, whereas the incident field ${\bf A}^i(t,{\bf r})$
 is considered as the external force for  scattering.

 Eliminating the
$z-$component, we write the Maxwell equation for the
$p-$polarization of the scattered field  ($x-z$ is the scattering
plane) in the form
%\begin{equation}\label{eqs1}
$$\left(\frac{d^2}{dz^2}
+k_z^{s2}(\omega^s)\right)E_x^s(\omega^s,k_x^s,z)=
I(\omega^s,k^s_x,z),$$
%\end{equation}
where
\begin{equation}\label{ci}
 I(\omega^s,k^s_x,z)=\frac{-4\pi
i}{\omega^s\varepsilon(\omega^s)}
\left(k_z^{s2}(\omega^s)j_x^s(\omega^s,k^s_x,z)+ik^s_x
\frac{dj_z^s(\omega^s,k^s_x,z)}{dz}\right)
\end{equation}
and $k_z^{s}(\omega^s)=
\sqrt{\varepsilon(\omega^s)(\omega^{s}/c)^2-k^{s2}_x}$ is the
normal component of the wave vector in the medium for the
scattered wave.

  The scattered field in the vacuum $z<z_0$ is expressed in terms of
$I(\omega^s,k^s_x,z)$:
%\begin{eqnarray}\label{fxz}
$$E_x^s(\omega^s,k^s_x,z)=
\frac{2\varepsilon(\omega^s)\omega^s\cos{\theta^s}/c}
{k_z^{s}+\varepsilon(\omega^s)\omega^s\cos{\theta^s}/c}
I_0e^{-i(z-z_0)\omega^s\cos{\theta^s}},$$
$$E^s_z(\omega^s,k^s_x,z)=(k^s\omega^s\cos{\theta^s}/ck^{s2}_z)
E^s_x(\omega^s,k^s_x,z),$$
%\end{eqnarray}
where
\begin{equation}\label{i0}
I_0=\frac{i}{2k^s_z}\int_{z_0}^{\infty}dz'
 e^{ik^s_z(z'-z_0)\cos{\theta^s}/c}I(\omega^s,k^s_x,z')
\end{equation}
and $\theta^s$ is the propagation angle of scattered wave in the
vacuum.

%\begin{equation}\label{ci}
% I(\omega^s,k^s,z)=\frac{4\pi
%i}{\omega^s\varepsilon(\omega^s)}
%\left(k^{s2}_zj_x^i+ik^s\frac{dj_z^i}{dz}\right),
%\end{equation}

The energy flow from the surface of the sample is given by
$|E_x^s(\omega_s,k^s_x,z<z_0)|^2$. Therefore, we have to calculate
\begin{equation}\label{cor}
<I^*(\omega^s,k^s_x,z)I(\omega^s,k^s_x,z')>
\end{equation}
averaged quantum-mechanically and statistically, where according
to Eqs. (\ref{ci}) and (\ref{cs}) we meet the Fourier transform of
the correlation function
%\begin{equation}\label{cor1}
$$K_{jk,j'k'}(t,{\bf r};t',{\bf r'})= <{\cal N}_{jk}^*(t,{\bf
r}){\cal N}_{j'k'}(t',{\bf r'})>.$$
%\end{equation}
 Because this correlator depends on the differences
$(t-t')$ and  $(x-x')$, we can expand that in the Fourier integral
with respect to these differences. Then, we have the Fourier
transform $K_{jk,j'k'}(\omega,k_x,z,z')$ which can be expressed
 in terms of the generalized susceptibility according to
the fluctuation-dissipation theorem:
$$K_{jk,j'k'}(\omega,k_x,z,z')=\frac{2}{1-\exp(-\omega/T)}
{\rm Im}\, \chi_{jk,j'k'} (\omega,k_x,z,z').$$

The generalized susceptibility $\chi_{jk,j'k'} (\omega,k_x,z,z')$
is involved in the response
\begin{equation}\label{gs}
<{\cal N}_{jk}(\omega,k_x,z)>=-\int_{z_0}^{\infty} dz'
\chi_{jk,j'k'} (\omega,k_x,z,z')U_{j'k'}(\omega,k_x,z')
\end{equation}
to the force
$$U_{jk}(\omega=\omega^i-\omega^s,k_x=k^i_x-k^s_x,z)=
A^s_j(\omega^s,k^s_x,z)A^i_k(\omega^i,k^i_x,z) \sim e^{iq_zz},$$
where $q_z=k_z^i+k_z^s$.

 To calculate the generalized
susceptibility $\chi_{jk,j'k'} (\omega,k_x,z,z')$, we
  write the equations for the averaged  phonon ${\bf u}$  and plasmon ${\bf E}$ fields.
  The variation of the Hamiltonian
(\ref{hami}) with respect to the vibrations ${\bf u}$ gives in the
right-hand side of the motion equation
 for the phonon field
\begin{equation}\label{ue1}
(\omega^2_{\rm TO}-\omega^2-i\omega\Gamma)u_i(\omega,k_x,z)=
\frac{Z}{\rho}E_i(\omega,k_x,z)-\frac{g^u_{ijk}}{\rho}
U_{jk}(\omega,k_x,z),
\end{equation}
  the additional term   to the force from the
electric field; $\rho$ is the density of the reduced mass and $Z$
is the effective charge.

The equation for the plasmon field ${\bf E}$ can be obtained if
this field is taken as the variable in the Hamiltonian
(\ref{hami}):
\begin{equation}\label{upf}
\nabla \cdot ({\bf E}+4\pi{\bf P})=4\pi\rho.
\end{equation}
The charge density $\rho$ is connected to the current by Eq.
(\ref{dens}). The polarization ${\bf P}$ includes   the dipole
moment $Z{\bf u}$, the contribution of the filled electronic state
$\chi_{\infty}{\bf E}$, and the variational term $-\delta {\cal
H}/\delta {\bf E}$:
%\begin{equation}\label{pol}
$$P_i=Zu_i+\chi_{\infty}E_i-g^{E}_{ijk}U_{jk}.$$
%\end{equation}
Here we can substitute $u_i$ using Eq. (\ref{ue1}). We obtain in
Eq. (\ref{upf}) the term with $\varepsilon_{\infty}=1+4\pi
\chi_{\infty}$ and $g^{E}_{ijk}\rightarrow {\tilde g}^{E}_{ijk}$.
\begin{equation}\label{tg}
{\tilde g}^{E}_{ijk}=
g^{E}_{ijk}+g^{u}_{ijk}Z/\rho(\omega^2_{\text
TO}-\omega^2-i\omega\Gamma).
\end{equation}

  Equation (\ref{upf}) can be substantially simplified in the case under
consideration when the the wave vectors ${\bf k}^i$, ${\bf k}^s$,
and consequently the momentum transfer ${\bf k}={\bf k}^i-{\bf
k}^s$ are determined by the frequency of the incident optical
radiation, whereas the frequency transfer $\omega$ of the excited
fields ${\bf E}$ and ${\bf u}$ is much less than the incident
frequency, $\omega=\omega^i-\omega^s\ll\omega^i$. Therefore,
  we can neglect here the retardation terms $\omega/c$ compared to
$k_x$ (so that $\kappa=k_x$) and introduce the potential, ${\bf
E}=- \nabla \phi$. We obtain for the potential the following
equation
\begin{equation}\label{phi}
\left(\frac{d^2}{dz^2}-k_x^2-2k_x
C\sum\delta(z-z_l)\right)\phi(\omega,k_x,z)=\frac{-4\pi}{\varepsilon(\omega)}
\left(ik_x{\tilde g}_{xjk}^E+{\tilde
g}^{E}_{zjk}\frac{d}{dz}\right)U_{jk}(\omega,k_x,z).
\end{equation}

The solution to this equation is found by using the Green's
function obeying the same Eq. (\ref{phi}) but with the
$\delta(z-z')$ function on the right-hand side. As very well
known, the Green's function is expressed in terms of the solutions
of the corresponding homogeneous equation. Taking into account
$f_{\pm}(\omega, k_x, z)$ (\ref{ef}), we write:
\begin{equation}\label{gf}
G(z,z')=\frac{i}{2\kappa\sin{(k_zd)}\sinh{(\kappa d)} } \left\{
\begin{array}{c}
f_{+}(z)f_{-}(z'), z>z' \\ \nonumber
f_{-}(z)f_{+}(z'), z<z'
\end{array}
\right .
\end{equation}
where where the quasi-momentum   $k_z=k_z(\omega^s,\kappa)$ has to
be determined from the dispersion relation (\ref{diseq}) ignoring
the retardation, $\kappa=k_x$.

According to
 $T_d$ symmetry of GaAs lattice, the Raman tensors $g^u_{ijk}$ and $g^E_{ijk}$
 have only two independent components $g_{xxx}$ and $g_{xyz}$ in the
 crystal axes. Let the  scattered light
is always polarized in the $xz-$plane. Now we consider two
geometry.  For the parallel scattering geometry (a), the incident
 light considered to be polarized in the $x-$ direction,
 then  the $x-$components of  exited phonon
 and plasmon fields can be active in the Raman scattering due to
 $g_{xxx}$. For the crossed geometry (b),  the incident light
 is polarized in the $y-$ direction. Then,   $z-$components of the exited
fields are active and we take into account terms with $g_{xyz}$.
Thus, for the generalized susceptibility
$\chi_{xx,xx}(\omega,k_x,z,z')$ defined by Eq. (\ref{gs}), we
obtain
\begin{equation}\label{xx}
\chi_{xx,xx}(\omega,k_x,z,z')=\frac{g^{u2}_{xxx}/\rho}{\omega^2_{\text
TO}-\omega^2-i\omega\Gamma}\delta(z-z')-\frac{4\pi
k_x^2}{\varepsilon(\omega)}\tilde{g}_{xxx}^EG(z,z')
\end{equation}

%To find the Raman cross-section, the susceptibility have to be
%integrated with the factor
%$U_{jk}^*(\omega,k_x,z)U_{jk}(\omega,k_x,z')$ over $z$ and $z'$,
%corresponding to Eqs. (\ref{fxz})--(\ref{cor}).

To get the Raman cross-section,  Eq. (\ref{cor}),
 we  evaluate the integral
%\begin{equation}\label{rc}
$$\int_{z_0}^{\infty}dz dz'e^{i(q_zz'-q_z^*z)}{\rm Im}\,
\chi_{ij,ij}(k,\omega,z,z').$$
%\end{equation}
We obtain for terms with $G(z,z')$
%\begin{eqnarray}\label{integ}
$$Int=\int_{z_0}^{\infty}dz f_{-}(z) \int_{z}^{\infty}dz'f_{+}(z')
e^{i(q_zz'-q_z^*z)}+ \int_{z_0}^{\infty}dz f_{+}(z)
\int^{z}_{z_0}dz'f_{-}(z') e^{i(q_zz'-q_z^*z)},$$
%\end{eqnarray}
and
\begin{eqnarray}\label{integ1}\nonumber
\int_{z_0}^{\infty}dz f_{-}(z) \int_{z}^{\infty}dz'f_{+}(z')
e^{i(q_zz'-q_z^*z)}=\\
\sum_{n,m=0}^{\infty}e^{i(q_z-q_z^*)nd+i(k_z-q_z)md}
\int_{z_0}^{z_0+d}
dzf_{-}(z)\int_{z}^{z+d}dz'f_{+}(z')e^{i(q_zz'-q_z^*z)}.
\end{eqnarray}
The sum over the integer $n$ could be extended to the infinity
being equal to $\delta/2d$, if the thickness of the sample is
larger compared with the skin-depth $\delta$ of the incident or
scattered waves. The sum over  $m$ reduces to a  not vanishing
factor only under the Bragg condition, $k_z-q_z=2\pi n/d$,
%gives
%$$Int=\frac{1}{2}(\delta/d)2\Delta(q_z-q),$$ where $\delta$ is the
%skin-depth of the incident or scattered waves. The function
%$\Delta(q_z-q)$
which expresses  the momentum conservation law  in the exciting
processes of the phonon-plasmon coupled modes.
%: $\Delta(q)=1$, if
%${\rm Re}\, q=2\pi n/d$ with the integer $n$
% and $\Delta(q)=0$ elsewhere.
In the macroscopic limit,
%$k, q, \kappa<<1/d$
 when the wave length
of the exited mode is large compared to the period $d$, only the
main Bragg maximum ($n=0$)  is observed for each of the coupled
modes. In this case, we can expand in powers of $k_xd$ and $k_zd$
the matrix element in Eq. (\ref{integ1}).
%$$Int=\frac{1}{2}(\kappa d\delta)2\Delta(q_z-q).$$
%The integral corresponding to $G_z(z,z')$ contains an additional
%factor $-iq_z.$

Omitting the common factors, we have  the Raman intensity for the
parallel geometry (a) in the form
\begin{eqnarray}
\label{ramx} Int_{xx} (\omega,k_x)={\rm
Im}\,\left\{\frac{g^{u2}_{xxx}/\rho} {\omega_{\rm
TO}^2-\omega^2-i\omega\Gamma}\right.+\\ \nonumber \left(g_{xxx}^E+
\frac{g_{xxx}^uZ/\rho}{\omega^2_{\rm
TO}-\omega^2-i\omega\Gamma}\right)^2\left. \frac{4\pi
k^2_x}{(k_z^2-q_z^2)\varepsilon(\omega)} \right\}
\end{eqnarray}
and for the crossed geometry (b)
\begin{eqnarray}
\label{ramz} Int_{xy} (\omega,k_x)={\rm
Im}\,\left\{\frac{g^{u2}_{xyz}/\rho} {\omega_{\rm
TO}^2-\omega^2-i\omega\Gamma}\right.+\\ \nonumber \left(g_{xyz}^E+
\frac{g_{xyz}^uZ/\rho}{\omega^2_{\rm
TO}-\omega^2-i\omega\Gamma}\right)^2\left. \frac{4\pi
q^2_z}{(k_z^2-q_z^2)\varepsilon(\omega)} \right\}.
\end{eqnarray}

 The wave-vector $k_z$ of the coupled phonon-plasmon
mode  is determined by Eq. (\ref{diseq}). For example, if the
incidence is normal to the sample surface and $\theta$ is the
scattering angle, then $k_x=\omega^i\sin{\theta}/c$ and
$q_z=(\sqrt{\varepsilon(\omega^i)}+
\sqrt{\varepsilon(\omega^i)-\sin^2{\theta}})\omega^i/c$, where we
take into account that $\omega^i\approx\omega^s$.

\begin{figure}[h]
\resizebox{.50\textwidth}{!}{\includegraphics{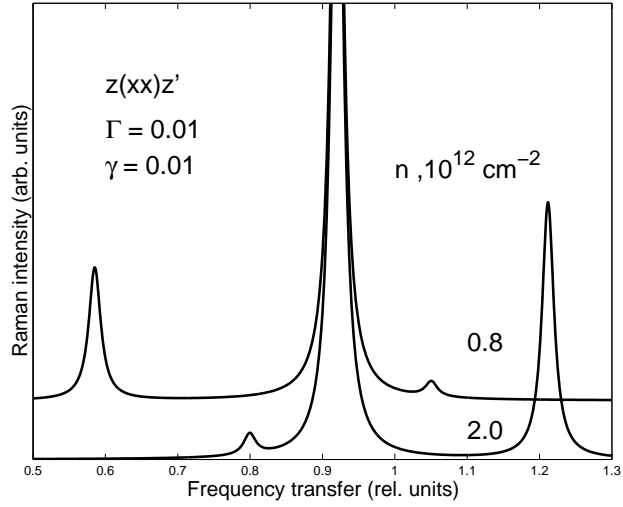}}
\caption{\label{pt} Raman intensity vs. the frequency transfer (in
units $\omega_{\rm LO}$) in the $z(xx)z'$  geometry for two values
of the current concentration indicated at the curves in units of
$10^{12}$ cm$^{-2}$; $z$ and $z'$ are the propagation directions
of the incident and scattered lights, respectively; the direction
$z'$ makes an angle of $\pi/4$ with the backscattering direction
$-z$, $.(xx).$ are the corresponding  polarizations. The values of
the other parameters are the same as in Fig. \ref{phpl1}. }
\end{figure}

\begin{figure}[h]
\resizebox{.50\textwidth}{!}{\includegraphics{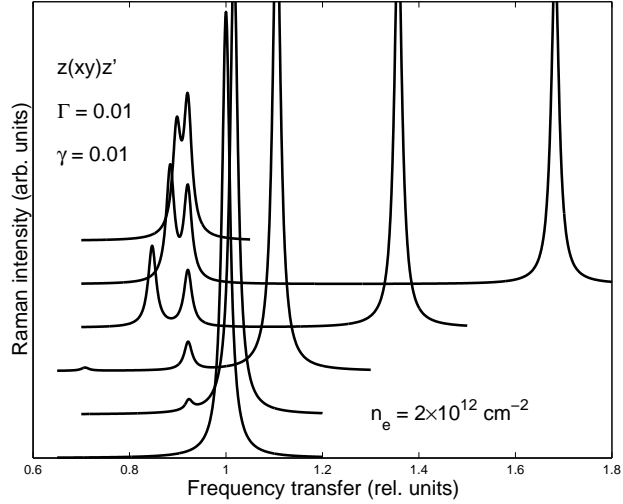}}
\caption{\label{prl}  Raman intensity vs. the frequency transfer
(in units $\omega_{\rm LO}$) in the $z(xy)z'$  geometry for the
angle of scattering varying from  0 (bottom) до $\pi/2$ (top) with
a step of $\pi/10.$}
\end{figure}

 In Eqs. (\ref{ramx}) -- (\ref{ramz}), we dropped slowly varying factors
 depending
 on the parameters of the incident and scattered radiations, for
 instance, their dielectric constants as well as the temperature factor
$1/[1+\exp{(-\omega/T)}].$
%The Raman intensity for geometry
% (b) differs by substitution of  $q_z$ for $k_x$ and vertices $g_{xyz}$
 %for $g_{xxx}$.
 %It is shown in Fig.
%\ref{prl} for various values of the scattering angle from 0
%(bottom) to $\pi/2$ (top) at an interval $\pi/10.$
In numerical calculations, we used the relationship between the
vertices $g^u$ and $g^E$ known from experiments  and given by the
Faust-Henry constant  $C_{FH}=g^uZ/g^E\rho\omega_{\rm TO}^2=-0.5.$
The results of calculations are shown in  Figs.  \ref{pt} and
\ref{prl}.

Note, that for the case of normal propagation of both incident and
scattered radiation, $k_x=0$, the second term in Eq. (\ref{ramx})
vanishes, and the Raman peak at parallel polarizations  (geometry
a) is situated at the frequency $\omega_{\rm TO}$. The others
peaks in Fig. \ref{pt} correspond to the excitations of the
coupled phonon-plasmons. But for the crossed polarizations
(geometry b) and $k_x\rightarrow 0$, the dispersion equation
(\ref{diseq}) with $C\rightarrow 0$ gives $k_z=ik_x$. Then, using
Eq. (\ref{ramz}) and the relationship $\omega_{\rm
LO}^2-\omega_{\rm TO}^2=4\pi Z^2/\varepsilon_{\infty}\rho$ between
the frequencies of the longitudinal and transverse phonons, we see
that the peak appears only at $\omega_{\rm LO}$, since the terms
with poles at $\omega_{\rm TO}$ are cancelled. This peak
corresponds to the zero of dielectric constant
$\varepsilon(\omega)$. At other scattering angles, a peak appears
at $\omega_{\rm TO}$ independently of the scattering angle.  Two
other peaks on each curves in Fig. \ref{prl} correspond to the
excitation of the phonon-plasmons.

\section{Conclusions}
In this work, we investigated  the infrared absorbtion and the
Raman scattering on the coupled phonon-plasmon modes with the help
of a simple model of superlattices formed by thin conducting
layers separated with insulating layers. This model admits the
dispersion relation of an analytical form. Our results for  the
reflectance and the Raman spectra show that the observed picture
can be drastically modified by means of the carrier concentration,
the superlattice period, and the frequency.

\begin{acknowledgments}
 One of us (L.A.F.) is grateful to J. Camassel for collaboration in
GES (Montpellier, CNRS) and P. Fulde (Dresden, MPIPKS) for
hospitality. The work was supported by the Russian Foundation for
Basic Research, project no. 04-02-17087.
\end{acknowledgments}

\end{document}